\documentclass[aps,prx,twocolumn,superscriptaddress,nofootinbib,showkeys]{revtex4-1}

\usepackage{graphicx}
\usepackage{dcolumn}
\usepackage{bm}
\usepackage[utf8]{inputenc}
\usepackage{textcomp}
\usepackage{amsmath}
\usepackage[mathlines]{lineno}
\usepackage{siunitx}
\usepackage{pdfpages}
\usepackage{pgffor}
\usepackage{pax}
\usepackage{hyperref}
\usepackage{placeins}

\hypersetup{
    colorlinks=true,
    citecolor=blue,
    urlcolor=blue,
    linkcolor=black
}

\makeatletter
\AtBeginDocument{\let\LS@rot\@undefined}					
\makeatother

\begin{document}

\title{Coupling interlayer excitons to whispering gallery modes in van der Waals heterostructures}

\author{Ronja Khelifa}
\affiliation{Photonics Laboratory, ETH Zürich, 8093 Zürich, Switzerland }
\author{Patrick Back}
\affiliation{Photonics Laboratory, ETH Zürich, 8093 Zürich, Switzerland }
\author{Nikolaus Fl\"ory}
\affiliation{Photonics Laboratory, ETH Zürich, 8093 Zürich, Switzerland }
\author{Shadi Nashashibi}
\affiliation{Photonics Laboratory, ETH Zürich, 8093 Zürich, Switzerland }
\affiliation{Present address: Institute of Electromagnetic Fields, ETH Zürich, 8092 Zürich, Switzerland}
\author{Konstantin Malchow}
\affiliation{Photonics Laboratory, ETH Zürich, 8093 Zürich, Switzerland }
\affiliation{Present address: Laboratoire Interdisciplinaire Carnot de Bourgogne, UMR 6303 CNRS-Université de Bourgogne Franche-Comté, 21078 Dijon, France}
\author{Takashi Taniguchi}
\affiliation{International Center for Materials Nanoarchitectonics, National Institute for Materials Science, 1-1 Namiki, Tsukuba 305-0044, Japan}
\author{Kenji Watanabe}
\affiliation{Research Center for Functional Materials, National Institute for Materials Science, 1-1 Namiki, Tsukuba 305-0044, Japan}
\author{Achint Jain}
\affiliation{Photonics Laboratory, ETH Zürich, 8093 Zürich, Switzerland }
\author{Lukas Novotny}
\affiliation{Photonics Laboratory, ETH Zürich, 8093 Zürich, Switzerland }
\date{\today}

\begin{abstract}
Van der Waals heterostructures assembled from two-dimensional materials offer a promising platform to engineer structures with desired optoelectronic characteristics. Here we use waveguide-coupled disk resonators made of hexagonal boron nitride (h-BN) to demonstrate cavity-coupled emission from interlayer excitons of a heterobilayer of two monolayer transition metal dichalcogenides. We sandwich a MoSe\textsubscript{2} -- WSe\textsubscript{2} heterobilayer between two slabs of h-BN and directly pattern the resulting stack into waveguide-coupled disk resonators. This enables us to position the active materials into regions of highest optical field intensity, thereby maximizing the mode overlap and the coupling strength. Since the interlayer exciton emission energy is lower than the optical band gaps of the individual monolayers and since the interlayer transition itself has a weak oscillator strength, the circulating light is only weakly reabsorbed, which results in an unaffected quality factor. Our devices are fully waveguide-coupled and represent a promising platform for on-chip van der Waals photonics.
\end{abstract}


\maketitle
Van der Waals (vdW) heterostructures made of a vertical assembly of two-dimensional (2D) materials like transition metal dichalcogenides (TMDCs) and graphene have opened up new possibilities for optoelectronic devices \cite{Bie2017, Flory2020}.
Since adjacent 2D materials are held together by weak van der Waals forces, they can be stacked without need of lattice matching and can be placed on almost any substrate \cite{Geim2013, Liu2016}.
In their monolayer form, TMDCs exhibit a direct bandgap \cite{Splendiani2010, Mak2010}, which makes them excellent candidates for optoelectronic devices as well as for integration with optical cavities for enhanced light matter interactions \cite{Wu2015, Ye2015, Reed2016, Li2017, Shang2017, Javerzac-Galy2018}.\par
By combining two dissimilar TMDC monolayers into a heterobilayer (HBL), interlayer excitons with spatially separated electrons and holes can be formed \cite{Rivera2015}.
Their charge carrier dynamics \cite{Rigosi2015}, electrically tunable exciton resonances \cite{Rivera2015, Jauregui2019} and the possibility of creating p-n junctions \cite{Ross2017}, are providing intriguing possibilities for materials engineering. 
Recently, interlayer excitons were coupled to photonic crystal cavities \cite{Liu2019, Rivera2020}, as well as grating cavities \cite{Paik2019}, that led to the demonstration of surface emitting, optically pumped lasers at room-temperature \cite{Liu2019, Paik2019}.

\par
\begin{figure*}[t!]
\centering
\includegraphics[width=1\textwidth]{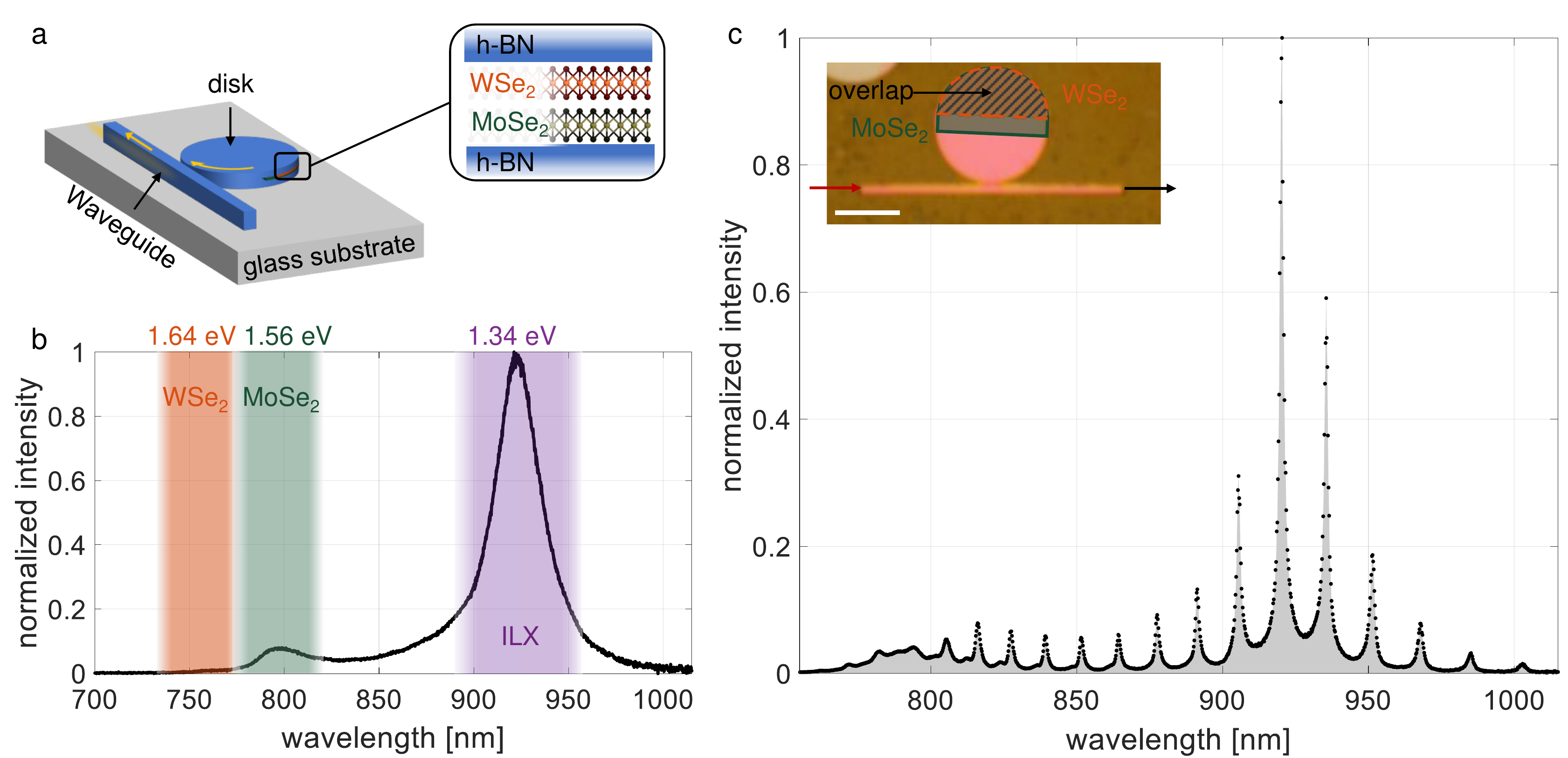}
\caption{(a) Illustration of a waveguide-coupled disk resonator with integrated heterobilayer (HBL). The HBL is sandwiched between two h-BN flakes. The emission from the interlayer exciton is coupled to whispering gallery modes (WGM) of the disk resonator and then outcoupled through the same waveguide used for excitation. (b) Free-space photoluminescence (PL) spectrum of the HBL, showing strong emission from the interlayer exciton (ILX) and quenched intralayer exciton emission. (c) Cavity-coupled photoluminescence spectrum featuring distinct peaks separated by the free-spectral range of the WGMs of the disk resonator. The resonator is excited at $\lambda=\SI{633}{\nano \meter}$ from one end of the waveguide and the photoluminescence is detected at the other end. The intralayer exciton emission is strongly suppressed because of reabsorption. The scale bar is \SI{5}{\micro \meter}.}  \label{fig:Schematic}
\end{figure*}\par
However, the co-integration of light emitting devices (LEDs) and lasers made of vdW heterostructures with on-chip photonic structures, such as waveguides and resonators, requires in-plane light guiding. 
Coupling via evanescent fields has been achieved, for example, by placing vdW materials on top of silicon-based  waveguides \cite{Bie2017, Flory2020} and ring resonators \cite{Wei2015}, but the coupling strength obtainable in this manner is modest because the highest field intensity is found inside the light confining dielectric medium.
Optimal mode overlap can be achieved by integrating the active layer into dielectric waveguides or resonators. 
For optically excited systems, this has been shown by sandwiching a monolayer WS\textsubscript{2} into a microdisk made of Si$_3$N$_4$/hydrogen silsesquioxane (HSQ) \cite{Ye2015} or hexagonal boron nitride (h-BN)  \cite{Ren2018}. 
However, in both of these structures the disk was freestanding and is not easily compatible with on-chip photonics.\par
Here we present vdW heterostructures patterned into waveguide-coupled disk resonators, as illustrated in Fig. \ref{fig:Schematic}a. 
The stacking allows us to place TMDCs into regions of high field intensity. 
Our approach combines two important attributes: 1) an enhanced optical mode overlap and 2) waveguide coupling for on-chip integration. 
These are enabling factors for the development of on-chip optoelectronic devices, such as LEDs, modulators and photodetectors.
For the dielectric material we chose h-BN, which has already been used to fabricate photonic structures like resonators or photonic crystals \cite{Kim2018, Ren2018, Froch2019} and has been proposed as a confinement layer in integrated photonics \cite{Ren2019}. 
In addition to its favorable photonic attributes, h-BN also provides a means of encapsulation, that is, by placing TMDCs in between h-BN we prevent them from degradation \cite{Lee2015}.\par 
As shown in Fig.~\ref{fig:Schematic}a, we integrate a HBL into an h-BN disk resonator, which can be excited by laser light propagating along a nearby waveguide.
Photoluminescence (PL) associated with the interlayer exciton is coupled out by the adjacent waveguide. 
The unique feature of our approach is that both, the active material (TMDCs) as well as the dielectric (h-BN) are assembled into a single vdW heterostructure and subsequently patterned into the final photonic circuit (disk resonator and waveguide). 
This differs from previous works \cite{Bie2017, Flory2020, Wei2015}, where TMDCs were transferred on top of pre-fabricated photonic elements.
As illustrated in Fig.~\ref{fig:Schematic}b, the interlayer exciton emission energy is lower than the optical band gaps of the individual monolayers, which leads to weak reabsorption of the interlayer emission. 
Reabsorption is further suppressed by the weak oscillator strength of the direct interlayer exciton transition \cite{Ross2017, Forg2019}. 
As a result, the quality (Q)-factor of the coupled mode is largely unaffected by reabsorption. 
Figure~\ref{fig:Schematic}c shows the spectrum of the waveguide-coupled interlayer emission. 
The vdW disk has been excited from one end of the waveguide with a \SI{633}{\nano \meter} laser and the PL has been detected at the other end. 
In the following we first describe the fabrication procedure, then present the characteristics of the HBL and h-BN disks and finally discuss the results of the combined system.\par
\begin{figure}[!ht] 
\centering
\includegraphics[width=0.8\columnwidth]{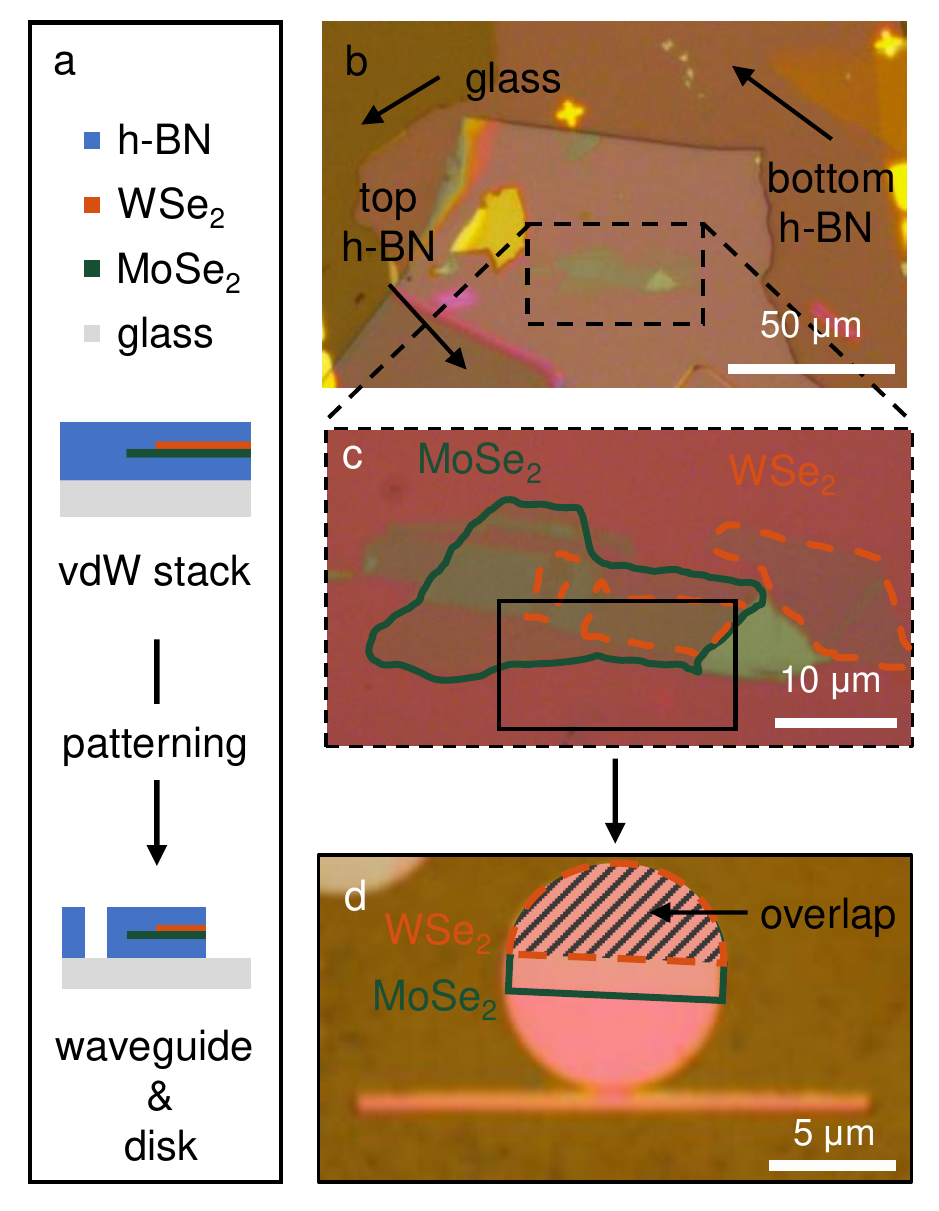}
\caption{(a) Schematic of the fabrication procedure, showing cross-sections of the van der Waals (vdW) heterostructure after stacking and after subsequent patterning into a waveguide-coupled disk resonator (not to scale). (b,c) Optical microscope images of the h-BN -- MoSe\textsubscript{2} -- WSe\textsubscript{2} -- h-BN heterostructure after stacking. The heterostructure has a total thickness of $\sim$\SI{310}{\nano \meter} (bottom h-BN: $\sim$\SI{150}{\nano \meter} and top h-BN: $\sim$\SI{158}{\nano \meter}). (d) Electron-beam lithography and reactive ion etching were utilized to pattern the vdW heterostructure into a waveguide-coupled disk resonator with a design radius of \SI{4.25}{\micro \meter}}  \label{fig:fabrication}
\end{figure}

Flakes of h-BN and monolayers of WSe\textsubscript{2} and MoSe\textsubscript{2} were prepared by mechanical exfoliation and assembled into vdW heterstructures using a polymer-based stacking technique \cite{Wang2013, Zomer2014}. 
The crystallographic axes of the two monolayers were rotationally aligned. 
Atomic force microscopy was utilized to identify h-BN flakes of the desired thicknesses. 
Figures~\ref{fig:fabrication}b and c (magnified) show optical microscope images of such an h-BN -- MoSe\textsubscript{2} -- WSe\textsubscript{2} -- h-BN heterostructure on a glass substrate.
As illustrated in Fig.~\ref{fig:fabrication}a, waveguides and disk resonators were fabricated from the vdW stack using electron-beam lithography and reactive ion etching with aluminum as a hard mask.
Fig.~\ref{fig:fabrication}d shows the final device.
The disk resonator has a total thickness of $\sim$\SI{310}{\nano \meter} and a design radius of \SI{4.25}{\micro \meter}. 
A more detailed description of the fabrication procedure can be found in the methods section in the Supporting Information (SI).\par
In order to determine the emission energies we performed PL measurements of the TMDC monolayers and of the HBL region using a \SI{633}{\nano \meter} continuous wave (cw) He-Ne laser.
The PL spectra of the individual monolayers reveal direct intralayer exciton emission peaks at \SI{755}{\nano \meter} (\SI{1.64}{\eV}) and \SI{793}{\nano \meter} (\SI{1.56}{\eV}) for WSe\textsubscript{2} and MoSe\textsubscript{2}, respectively, and are in agreement with previous works \cite{Ross2013, Rigosi2015, Karni2019, Liu2019} (details in the SI). 
In Fig.~\ref{fig:Schematic}b we show the PL spectrum of the HBL, with a pronounced peak at around \SI{923}{\nano \meter} (\SI{1.34}{\eV}).
This feature is identified as the interlayer exciton emission and can be explained by the type-II band alignment of MoSe\textsubscript{2} and WSe\textsubscript{2} in the overlap region \cite{Rivera2015, Paik2019}. 
In this process the absorption of a photon in a monolayer is followed by a charge transfer to the energetically favorable band of the neighboring monolayer, leading to a spatial separation of electron and hole.
As a result, a strong decrease of the intralayer exciton emissions can be observed.
Furthermore, for two rotationally aligned monolayers the interlayer exciton exhibits a direct band gap character \cite{Zhang2017}. 
The strong interlayer peak in combination with the quenching of intralayer exciton emission in Fig.~\ref{fig:Schematic}b consequently verifies the good electrical coupling and the near perfect rotational alignment of the two layers.\par
\begin{figure}[!ht] 
\centering
\includegraphics[width=0.8\columnwidth]{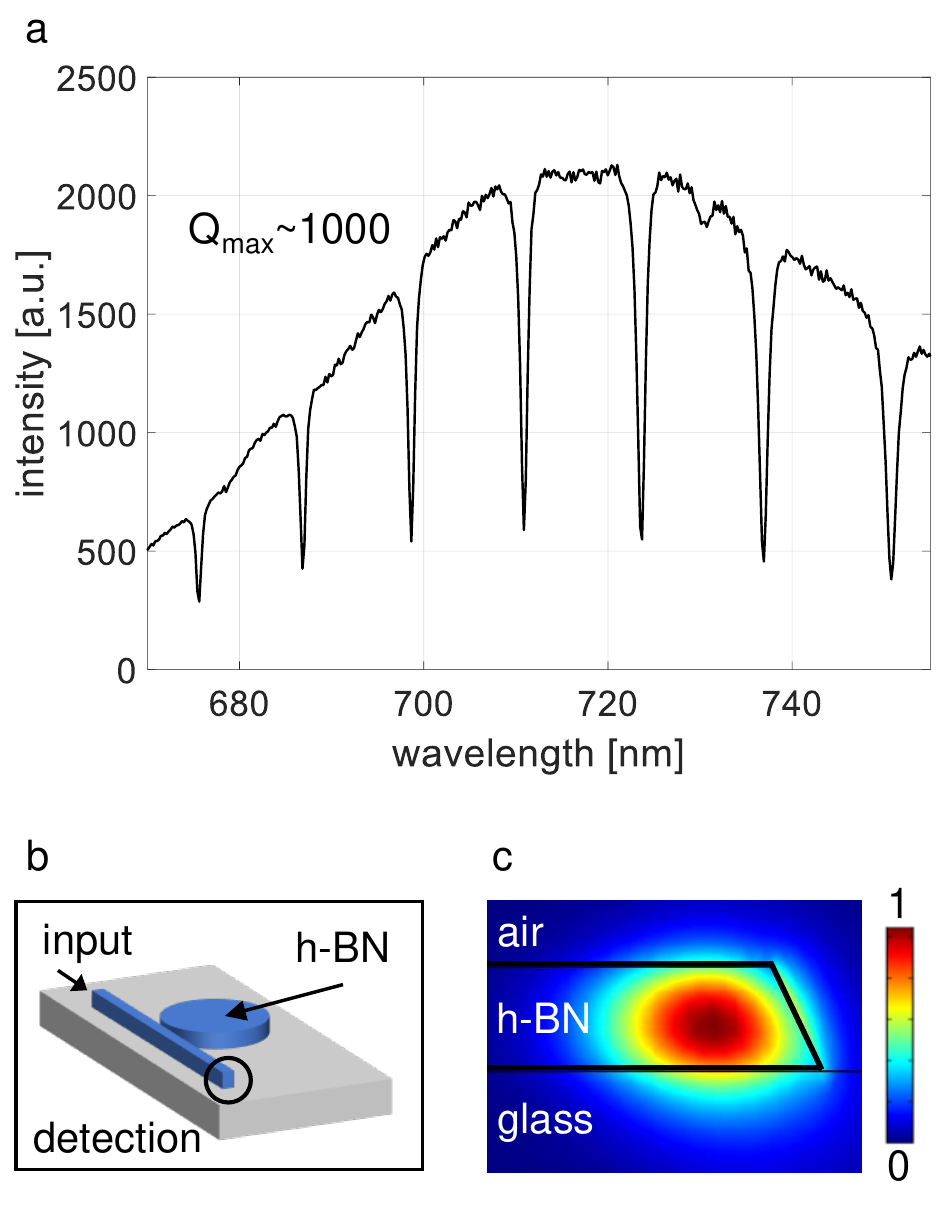}
\caption{(a) Transmission spectrum of an h-BN waveguide-coupled disk resonator without integrated TMDC. The dips correspond to the resonances of the WGMs and the envelope is determined by the spectrum of the broadband laser source and the transmission characteristics of the setup and waveguide. (b) Sketch of the h-BN waveguide-coupled disk resonator. (c) Simulated electric field distribution (absolute value) near the rim of a disk resonator (cross-sectional view).}  \label{fig:simulations}
\end{figure}

To investigate the properties of h-BN waveguide-coupled disk resonators we fabricated reference structures without integrated TMDCs. 
Figure~\ref{fig:simulations}a shows a typical transmission spectrum of a fabricated device.
It was obtained by focusing a broadband supercontinuum laser on one waveguide facet and detecting the transmitted light at the other end of the waveguide.
The recorded spectrum features sharp dips at wavelengths that satisfy the resonance conditions of the disk resonator.
The envelope of the spectrum is defined by the excitation spectrum and the transmission characteristics of the experiment. 
The Q-factors of the resonances reach values of $\sim$1000. 
The spectral positions of the resonance peaks, the Q-factors and the free spectral range depend sensitively on the h-BN thickness and the disk radius (details in the SI). 
Using finite element simulations as a guide, we define the disk radius for a given stack thickness in order to optimize the resonances for the interlayer emission spectrum (details on simulations in the SI). 
As shown in Fig.~\ref{fig:simulations}c, the computed electric field distribution is highest near the rim of the disks and near the vertical center. 
Therefore, to optimize the coupling strength to the interlayer exciton the HBL should be sandwiched into the h-BN resonator.\par
\begin{figure*}[!t]
\includegraphics[width=1\textwidth]{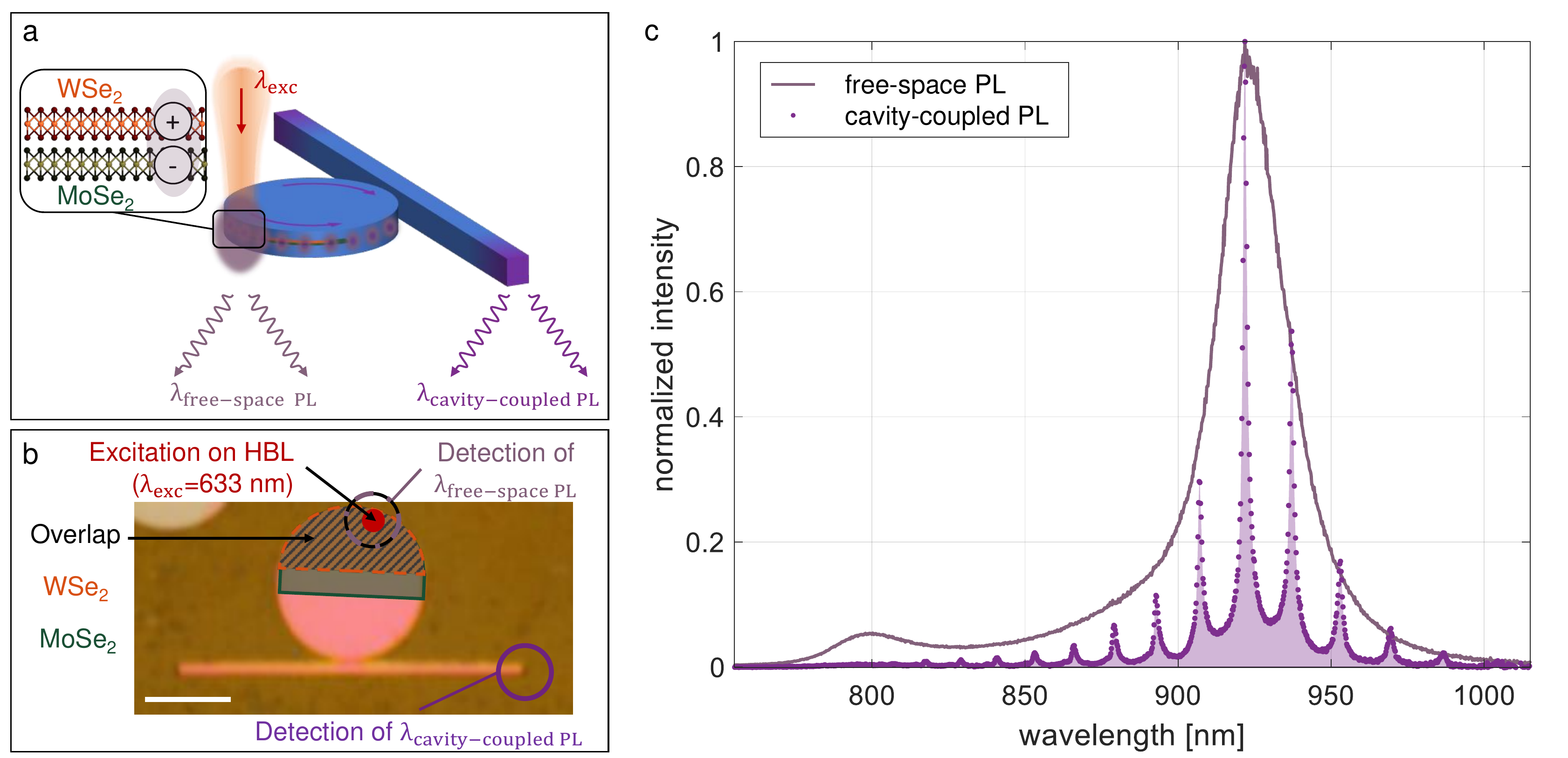}
\caption{(a) Illustration of cavity-coupled and free-space detection. Comparison of the spectra recorded by the two detection schemes provides information on reabsorption in the h-BN resonator.  (b) Optical microscope image indicating the locations of excitation and detection, and the regions with MoSe\textsubscript{2}, WSe\textsubscript{2} and their overlap (HBL region). The structure is excited with a \SI{633}{\nano \meter} cw laser at the edge of the disk, on the HBL region. The scale bar is \SI{5}{\micro \meter}. (c) Measured spectrum at the waveguide end (cavity-coupled PL) and at the excitation spot (free-space PL). Interlayer emission can be detected at the ends of the waveguide. The PL detected at the excitation spot coincides well with the envelope function of the cavity-coupled PL.}  \label{fig:PL_and_resonances}
\end{figure*}
Next, we characterize an h-BN disk resonator with integrated TMDC heterobilayer. 
We vary the location of the excitation in order to investigate the coupling of the interlayer emission to the WGMs of the disk. 
As shown in the sketch and the optical microscope image in Figs.~\ref{fig:PL_and_resonances}a and b, coupling to WGMs and interlayer excitons can be achieved by placing the excitation spot near the edge of the disk. 
We use the same objective for excitation and detection. 
The emission is detected either at the same position as the excitation spot or at the ends of the waveguide. 
The light coupled out at the end of the waveguide is partially refracted into the glass substrate and can be captured at high angles with the oil-immersion objective (NA=1.4) that we use. 
This allows us to compare the spectra of free-space PL (detected at the excitation spot on the disk) and cavity-coupled PL (detected at the waveguide end). 
As shown in Fig.~\ref{fig:PL_and_resonances}c, the spectrum of the free-space PL is in agreement with the expected interlayer emission of the HBL (cf. Fig.~\ref{fig:Schematic}b). 
On the other hand, the cavity-coupled spectrum, detected at the waveguide end,  exhibits distinct peaks that are separated by the free-spectral range of the WGMs of the disk resonator. 
We emphasize that, in contrast to several other recent studies \cite{Ye2015, Reed2016, Ren2018}, the waveguide-coupled spectrum features a reduced background because it is not superimposed to free-space PL.\par
\begin{figure}[!ht]
\includegraphics[width=1\columnwidth]{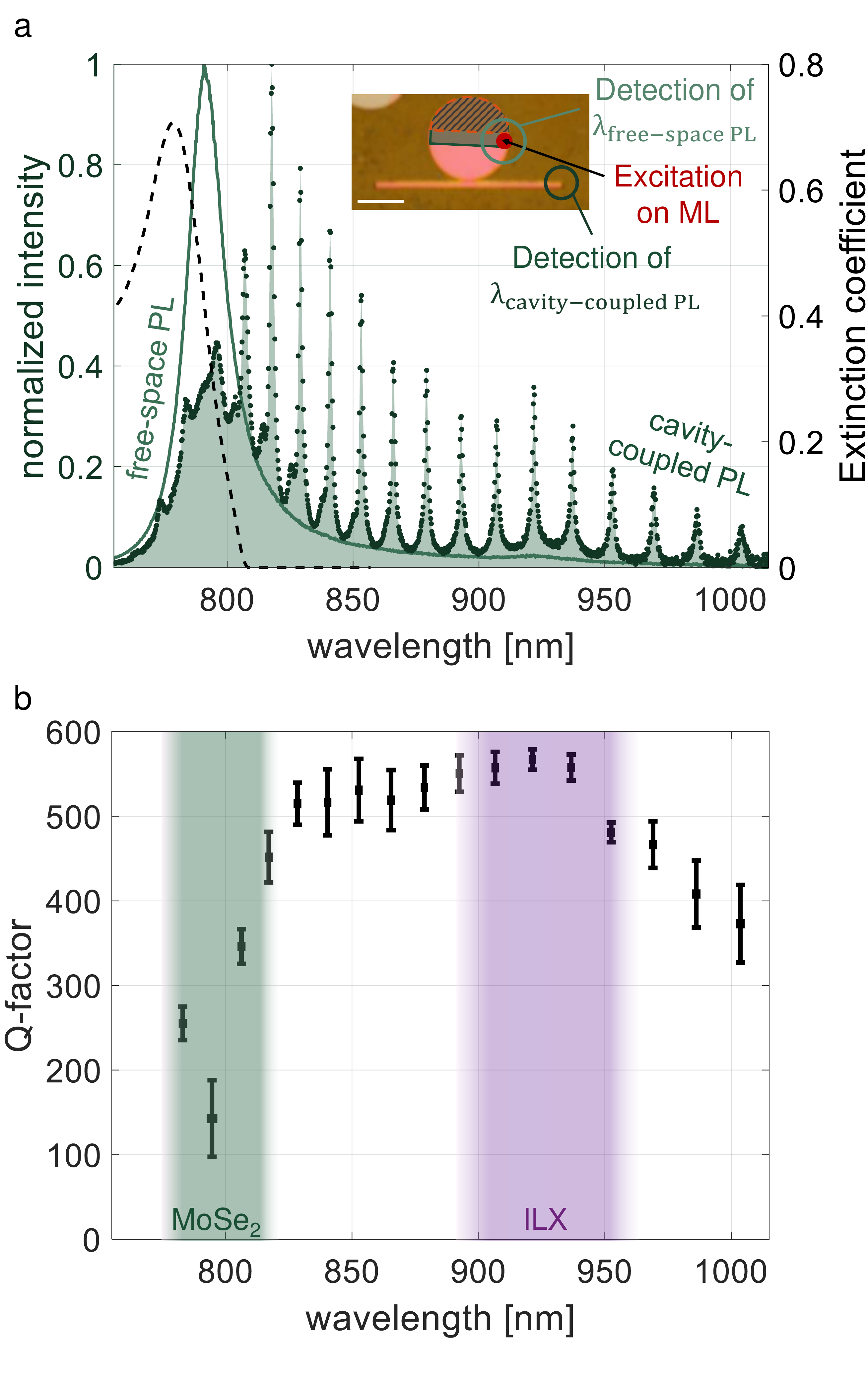}
\caption{ (a) Measured spectra at the waveguide end (cavity-coupled PL) and at the excitation spot (free-space PL), when exciting at the MoSe\textsubscript{2} monolayer region, as indicated in the inset (scale bar is \SI{5}{\micro \meter}). The envelope function of the cavity-coupled PL does not coincide with the free-space PL. Indicating strong reabsorption of intralayer emission by the MoSe\textsubscript{2} monolayer itself. The dashed curve indicates the MoSe\textsubscript{2} extinction spectrum (taken from \cite{Hsu2019}). (b) Wavelength dependence of the Q-factor, determined by fitting Lorentzian functions to the measured resonances. Measurements were taken for different locations of excitation (HBL, monolayer, waveguide end) and detection at both waveguide ends. The means and the standard deviations are shown in the figure. A strong drop in Q-factor can be observed at the absorption wavelengths of the intralayer exciton transition of MoSe\textsubscript{2}. No drop in Q-factor is observed for wavelengths of the interlayer exciton transition, indicating low reabsorption.}  \label{fig:Q-factor}
\end{figure}

As shown in the inset of Fig.~\ref{fig:Q-factor}a,  the two TMDCs forming the HBL do not overlap over the entire footprint of the disk. 
There are regions where only the  MoSe\textsubscript{2} monolayer can be directly excited. 
Figure~\ref{fig:Q-factor}a shows the corresponding spectrum of the free-space and cavity-coupled PL. 
The main contribution of the cavity-coupled PL comes from the intralayer exciton emission of the MoSe\textsubscript{2} monolayer.
Interestingly, however, the envelope function of the resonances does not coincide with the free-space PL, measured at the excitation spot.
Instead, the highest peak intensity of the cavity-coupled PL is positioned at wavelengths longer than the free-space PL emission peak.
 We attribute this observation mainly to reabsorption of the emitted PL in the cavity by the MoSe\textsubscript{2} monolayer itself, although other contributions have also been reported to be present \cite{Rosser2020a} (detailed discussion in the SI). 
This interpretation is supported by the absorption spectrum of MoSe\textsubscript{2}, visualized by the extinction coefficient, i.e. the imaginary part of the refractive index (taken from \cite{Hsu2019}), which we include in Fig.~\ref{fig:Q-factor}a (dashed curve).
Thus, the spectral overlap of the cavity-coupled PL with the extinction coefficient causes significant reabsorption for wavelengths in this spectral range. 
As a result, the corresponding peak intensities of the cavity-coupled PL are reduced for these wavelengths.
The peaks at longer wavelengths have an additional contribution from interlayer emission from the HBL. 
This emission is made possible by the reabsorption of the monolayer emission in the overlap region of the two TMDCs.
The effect of enhanced reabsorption in a sandwiched structure is further discussed in the Supporting Information, where two structures with integrated monolayer TMDCs are compared: 1) a sandwiched device and 2) a device with a monolayer transferred on top of the disk.
The comparison shows that reabsorption in a sandwiched device is significantly stronger, due to the enhanced mode overlap.\par
\par
The coupling of interlayer excitons to WGMs, profits not only from the enhanced mode overlap in the sandwiched structure, but also from low intrinsic reabsorption. 
The latter is due to the low oscillator strength of the interlayer transition, resulting from the spatial separation of the charge carriers \cite{Ross2017, Forg2019}. 
The oscillator strength of the interlayer exciton in MoSe\textsubscript{2} -- WSe\textsubscript{2} has been experimentally determined to be two orders of magnitude lower than that of the intralayer exciton \cite{Ross2017}.
In addition, in differential reflectance measurements, a reduced interlayer exciton absorption has also been shown, albeit at low-temperatures \cite{Forg2019}.
This is in agreement with our measurements shown in Fig.~\ref{fig:PL_and_resonances}c, where the HBL has been excited.
Here, the free-space PL agrees well with the envelope function of the cavity-coupled resonance peaks.
To summarize, our data indicates that the intralayer emission from monolayer TMDCs is strongly reabsorbed compared to the interlayer emission.
The integration of a HBL inside the disk resonator consequently has two benefits:  1) strong absorption of the excitation by intralayer excitons and 2) weak reabsorption of the interlayer emission within the cavity.\par
To allow for an absolute comparison of intralayer and interlayer coupling to WGMs, we now discuss waveguide-coupled excitation and read-out of the disk resonator (Fig.~\ref{fig:Schematic}c inset). 
The data in Fig.~\ref{fig:Schematic}c shows that the highest peak intensity is observed at wavelengths corresponding to the interlayer emission. 
In fact, the monolayer emission is strongly suppressed, in agreement with our previous conclusions.  
We estimate a ratio of $\sim$13 between the peak intensities of the main resonances ($\sim$\SI{920}{\nano \meter} and $\sim$\SI{820}{\nano \meter}) of interlayer and intralayer emission (details in the SI). 
Note that the spectrum in Fig.~\ref{fig:Schematic}c has been corrected for the wavelength dependent transmission of our measurement setup and the CCD detection efficiency. 
Our waveguide-coupled experiments support our previous conclusion that intralayer emission from TMDC monolayers is largely reabsorbed.\par
As shown in Fig.~\ref{fig:Q-factor}b, the Q-factors of the emission peaks are wavelength dependent, which is a result of the optical properties of the incorporated TMDC HBL.
The highest values are $\sim$570 and occur at wavelengths that correspond to the interlayer emission peak.
In contrast, a considerably lower Q-factor is observed for wavelengths near the optical band gap of MoSe\textsubscript{2}.
This provides further support for the reabsorption of monolayer emission discussed earlier. 
Because of the low oscillator strength of the interlayer transition the reabsorption of interlayer emission is significantly weaker, which is evidenced by the high Q-factor at wavelengths near the interlayer transition. 
The decrease in Q-factor at longer wavelengths ($>$\SI{950}{\nano \meter}) can be explained by radiation losses, which increase at higher wavelengths (see SI). 
Our Q-factors are currently limited by fabrication imperfections and the h-BN sidewall roughness arising from etching (see scanning electron microscope image in the SI). 
Nevertheless, even with integrated TMDCs and a coupled waveguide, we achieve Q-factors that are slightly higher than those reported for h-BN ring resonators \cite{Froch2019}. 
Our Q-factors could be increased even further by removing the substrate underneath the outer regions of the disk where the WGMs are localized \cite{Ye2015, Ren2018}.\par
In summary, we have demonstrated the coupling of interlayer excitons from a TMDC HBL to WGMs of an h-BN disk resonator. 
We achieve high coupling strengths by sandwiching the HBL into the h-BN disk. 
The high oscillator strengths of intralayer transitions of the two monolayers lead to a high absorption efficiency. 
On the other hand, the low oscillator strength of the interlayer transition suppresses the reabsorption at the emission wavelength. 
The combination of these two effects gives rise to strong interlayer emission from our vdW disk resonators.\par

The presented platform would give rise to further quantitative analysis of the cavity quantum electrodynamics and experimental studies on interlayer excitons, their dynamics and emission properties, when placed at high field intensities.
Considering the fact that patterning photonic structures out of h-BN and vdW heterostructures is still in its infancy, further improvements in performance with optimized fabrication techniques (e.g. optimization of the etching process) is expected.
To proceed one step further towards on-chip optoelectronics, light sources need to be co-integrated with photonic structures.
The recent progress in electrical control of interlayer excitons \cite{Jauregui2019}, as well as the realization of light emitting devices \cite{Ross2017,Binder2019,Jauregui2019}, demonstrates a wide range of possibilities for band structure engineering with HBLs.
Therefore, combining our approach with electrically tunable interlayer excitons, offers a promising platform for on-chip integrated photonics based on 2D materials.\par

\begin{acknowledgments}
This work has been supported by the Swiss National Science Foundation (grant 200020\_192362/1). 
The authors thank A.~Kuzmina, D.~Windey, S.~Papadopoulos, R.~Reimann, M.~Parzefall and S.~Heeg for fruitful discussions.
In addition the authors thank A.~\.{I}mamo\u{g}lu at ETH Zürich for the usage of the stacking setup. 
The use of the facilities of the FIRST center for micro- and nanoscience at ETH Zürich is gratefully acknowledged. 
K.W. and T.T. acknowledge support from the Elemental Strategy Initiative
conducted by the MEXT, Japan, Grant Number JPMXP0112101001,  JSPS
KAKENHI Grant Numbers JP20H00354 and the CREST(JPMJCR15F3), JST.
\end{acknowledgments}

\section*{Author contributions}
L.N. and R.K. conceived the project. R.K. and P.B. fabricated the devices and performed the experiments. A.J., N.F. contributed to the experiments. S.N. and K.M. performed preliminary studies. T.T. and K.W. synthesized the h-BN crystals. R.K., P.B., N.F., A.J. and L.N. analysed the data and co-wrote the manuscript.

\section*{Supporting Information}
The Supporting Information includes descriptions of: sample fabrication, methods of measurements, photoluminescence measurements, characterization of h-BN disk resonators, simulations of h-BN disk resonators, study of vertical emitter position inside h-BN disk resonator,  transmission measurement of waveguide-coupled disk resonator with integrated HBL, discussion of the effect of reabsorption inside the disk resonator,  determination of Q-factor, as well as comparison of the resonance peak intensities for the waveguide-coupled disk resonator with integrated heterobilayer.\par

\providecommand{\latin}[1]{#1}
\makeatletter
\providecommand{\doi}
  {\begingroup\let\do\@makeother\dospecials
  \catcode`\{=1 \catcode`\}=2 \doi@aux}
\providecommand{\doi@aux}[1]{\endgroup\texttt{#1}}
\makeatother
\providecommand*\mcitethebibliography{\thebibliography}
\csname @ifundefined\endcsname{endmcitethebibliography}
  {\let\endmcitethebibliography\endthebibliography}{}

\onecolumngrid 	



\section*{Supplementary material}

# Supporting Information: Coupling interlayer excitons to whispering gallery modes in van der Waals heterostructures

Ronja Khelifa,[1] Patrick Back,[1] Nikolaus Flöry,[1] Shadi Nashashibi,[1,2] Konstantin Malchow,[1,3] Takashi Taniguchi,[4] Kenji Watanabe,[5] Achint Jain,[1] and Lukas Novotny[1]

[1]*Photonics Laboratory, ETH Zürich, 8093 Zürich, Switzerland*
[2]*Present address: Institute of Electromagnetic Fields, ETH Zürich, 8092 Zürich, Switzerland*
[3]*Present address: Laboratoire Interdisciplinaire Carnot de Bourgogne, UMR 6303 CNRS-Université de Bourgogne Franche-Comté, 21078 Dijon, France*
[4]*International Center for Materials Nanoarchitectonics, National Institute for Materials Science, 1-1 Namiki, Tsukuba 305-0044, Japan*
[5]*Research Center for Functional Materials, National Institute for Materials Science, 1-1 Namiki, Tsukuba 305-0044, Japan*

## I. METHODS

### A. Sample Fabrication

Monolayer MoSe$_2$ and WSe$_2$ as well as multilayer hexagonal boron nitride (h-BN) flakes were mechanically exfoliated on top of Si/SiO$_2$ substrates. An optical microscope was used to identify suitable flakes and atomic force microscopy scans allowed the exact determination of the flake thicknesses. The heterostructure was stacked under inert argon (Ar) atmosphere, using a polydimethylsiloxane (PDMS) stamp, covered with a polycarbonate (PC) film. Care was taken to rotationally align the MoSe$_2$ and WSe$_2$ monolayers. The heterostructure was then transferred on top of a glass coverslip with gold markers. The structures were defined by means of electron-beam lithography using Polymethylmethacrylat (PMMA). The positioning and dimensions of the disk resonators were adapted to the shape of the individual flakes and the overall stack thickness. To prevent charging, an electrically conductive polymer, Espacer 300z was spin coated over the PMMA. Following the removal of the Espacer and the development of the PMMA resist, 50 nm aluminum was evaporated to serve as a hard mask. Reactive ion etching with SF$_6$ and Ar (20 sccm:20 sccm) was used to etch the stack with an RF power of 95 W and a chamber pressure of 100 mTorr. As a starting point for the etching we used parameters from [1]. The etching was performed in short time steps with a total time that depended on the sample thickness. For the presented heterostructure with integrated heterobilayer, the total etch time was adjusted to 40 sec. The Al mask was removed with a tetramethylammonium hydroxide based solution.

### B. Measurements

All measurements were performed in ambient conditions. For the transmission measurements in Fig. 3a a broad band pulsed laser was used. For the free-space and cavity-coupled photoluminescence (PL) measurements we used a 633 nm continuous wave (cw) He-Ne laser. The beam was focused with an oil-immersion objective (Nikon, 100x, NA 1.4) and was measured to be ∼47 µW for Fig. 4c, Fig. 5a and ∼186 µW for Fig. 1c before the objective. The sample was positioned on an x,y piezo stage, which allowed the precise positioning of the incoming laser light onto the waveguide or the region of interest on the disk. The excitation with normal incidence impedes the excitation with the electric field, polarized out-of-plane with respect to the sample plane. The emitted light was collected with the same objective from the entire structure and a 633 nm long pass filter was used to block the laser source. A Princeton Instruments Acton SpectraPro 300i spectrometer and a 300 grooves per mm grating were utilized for spatially resolved spectral measurements. The entrance slit width of the spectrometer was set to 50um. After dispersion by the grating, each row of the spectrometer CCD recorded a spectrum corresponding to a specific position along the slit. Imaging different parts of the device, i.e. excitation spot or waveguide facet, on the spectrometer opening allowed us to differentiate and compare spectra of individual regions. In a silicon CCD camera, the detection efficiency for longer wavelengths is strongly reduced. To take this into account, we corrected all measurements (except in Fig. 3a) to the transmission of the setup and the CCD camera. In addition, background spectra were detected and subtracted from all measurements.

## II. PHOTOLUMINESCENCE MEASUREMENTS

Figure S1 shows representative PL measurements of the heterostructure on the individual monolayers, as well as on the heterobilayer (HBL). On each region five measurements with a tightly focused (1.4 NA objective) cw 633 nm laser were executed and mean maximum peak positions were determined: WSe$_2$: $1.643\,\text{eV} \pm 0.002\,\text{eV}$, MoSe$_2$: $1.564\,\text{eV} \pm 0.001\,\text{eV}$ and for the interlayer exciton emission peak in the HBL region: $1.343\,\text{eV} \pm 0.002\,\text{eV}$. All measurements were taken under the same conditions and the spectra are corrected for the transmission of the setup and the detection efficiency of the CCD. This allows a comparison of the emitted intensity, which clearly shows that the two individual monolayer emit with higher intensities.

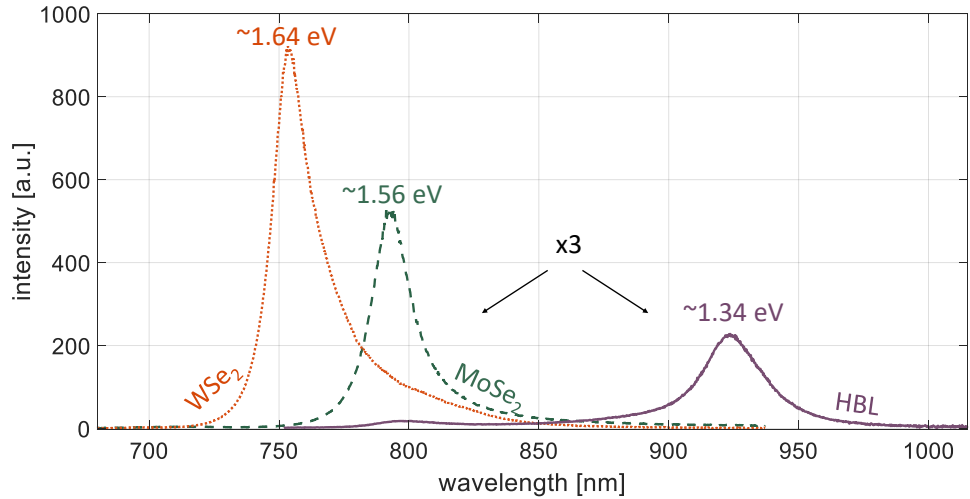

Fig. S 1. PL measurements on the individual monolayers WSe$_2$ and MoSe$_2$, as well as on the HBL region. The spectra were taken under the same conditions and are corrected for the transmission of the setup and the detection efficiency of the CCD.

## III. CHARACTERIZATION OF H-BN DISK RESONATORS

In the following we discuss in more detail the h-BN disk resonators without integrated TMDCs (schematic in Fig. S 2d). In Fig. S 2a the transmission measurement of the h-BN waveguide-coupled disk resonator, which was also discussed in Fig. 3a in the main text, is shown. The disk resonator has a radius of 3 µm and a height of ∼220 nm. To characterize the device, lorentzian lineshapes were fitted to the measured resonances and Q-factors of up to ∼1000 were determined (blue diamonds in Fig. S 2c). In general, several h-BN devices with different geometries and strongly varying Q-factors (∼200-1000) were fabricated. We believe the reason for the variations are the numerous properties that influence the Q-factor, e.g. the thickness of the h-BN flake. To visualize the influence of the height on the device performance, the transmission spectrum of another device with the same disk radius of 3 µm but a height of ∼175 nm is shown in Fig. S 2b and its corresponding Q-factor is plotted in Fig. S 2c (green circles). Different spectral resonance positions and a decreased Q-factor can be observed. Consequently, to develop an understanding on the most suitable geometries, we used simulations as a guide to optimize our devices and determine the best thicknesses and disk radii (discussed in section IV).
Overall, we believe the main factor that limits the Q-factors are fabrication imperfections, as for example the high scattering losses at the rough sidewalls, which are evident in the scanning electron microscopy (SEM) image of a representative device in Fig. 2e. The SEM image was taken directly after the dry etching step, before the metal mask was removed.

To further characterize the h-BN disk resonators, in the following we estimate the corresponding Purcell factor:

$$F_P = \frac{3Q}{4\pi^2 V_{eff}} \left(\frac{\lambda_0}{n}\right)^3 \tag{1}$$

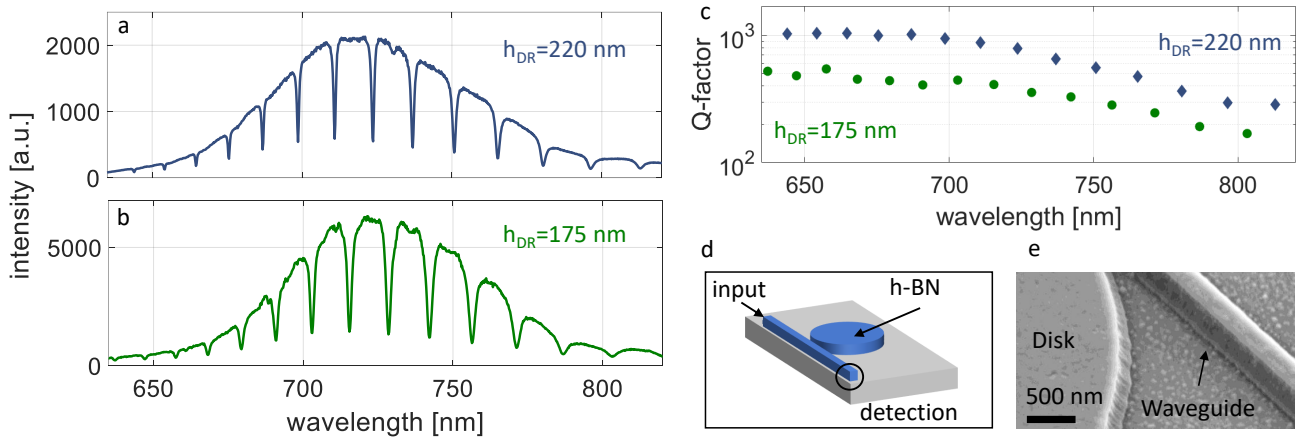

Fig. S 2. (a) Transmission measurement of an h-BN waveguide-coupled disk resonator with a radius of 3 μm and a height of ∼220 nm. (b) Transmission measurement of an h-BN waveguide-coupled disk resonator with the same radius but a height of ∼175 nm. (c) Q-factors determined from the measurements in (a) and (b). Even though the two devices have the same radius and were etched together for the same time, their Q-factor varies depending on their height. (d) Sketch of an h-BN disk resonator with indicated measurement scheme for the spectra in (a) and (b). (e), SEM image of a representative h-BN disk resonator directly after etching (aluminum mask is still on top).

where $V_{eff}$ is the effective mode volume, Q is the quality (Q)-factor and $\frac{\lambda_0}{n}$ is the resonance wavelength inside the material with the refractive index $n$. We estimate the enhancement by a cavity with a radius of 4.25 μm and a height of 310 nm for the resonance wavelength closest to 920 nm. We determine the effective mode volume from numerical simulations, whereas for the Q-factor we use a value of 570, which was determined from measurements. The estimated enhancement then comes out to be ∼3. The enhancement is not very strong owing to the relatively high effective mode volume and the modest Q-factor of the disk resonator in our case. In principle, a higher Purcell enhancement can for example be achieved by designing an optical cavity with a smaller effective mode volume.

For an accurate estimation of the Purcell enhancement of the interlayer exciton emission, the Q-factor of the emitter should also be considered [2, 3], this involves experimentally determining the radiative lifetime of the emitter which is beyond the scope of this study.

## IV. SIMULATIONS OF H-BN DISK RESONATORS

2D simulations of the disk resonator without coupled waveguide were executed using the wave-optics module of COMSOL multiphysics. For the simulations the refractive index of h-BN was defined as constant over the considered wavelength range and was defined as: $n_{\parallel} = 2.12$, $k_{\parallel} = 0.000048$, $n_{\perp} = 1.83$, $k_{\perp} = 0.000043$. The values for $k$ were taken from [4], whereas for $n$, we used values from [5] as an orientation. The h-BN disk was simulated on top of a glass substrate ($n = 1.52$) and surrounded by air ($n = 1$). Since the Q-factor of a resonance and the free spectral range depend amongst others on the disk radius and height, simulations were used to develop an understanding on the proper design parameters. Depending on the emission wavelength of the integrated emitter, the disk geometry was adjusted, according to the simulations. As an example Fig. S 3 shows the simulated Q-factors for devices with a height of 310 nm and a radius of 3 μm (red diamonds, and red dashed line as guide for the eye), as well as 4.25 μm (purple circles, and purple line as guide for the eye). According to these simulations, to integrate monolayer $MoS_2$, with an emission wavelength of ∼660 nm, as presented in section V, a radius of 3 μm is beneficial. However for this radius a strong decrease in Q-factor can be observed for higher wavelengths. Consequently, for the device with an integrated $MoSe_2$-$WSe_2$ HBL with the interlayer exciton emission at 923 nm, the designed disk radius should be adapted to a higher value. We choose 4.25 μm and not a higher radius, because with increasing disk radius the free spectral range between two adjacent resonances decreases (cf. Fig. S 3).

As expected, the simulated Q-factors are higher than the ones determined from measurements. We believe that the main reason for this are the fabrication imperfections, like for example the sidewall roughness. In addition, the simulations don't include the adjacent waveguide.

In general the decrease in Q-factor for increasing wavelengths is also observable in measurements: for the device with integrated heterobilayer and for h-BN devices without integrated emitter (cf. Fig. 5b in the main text and

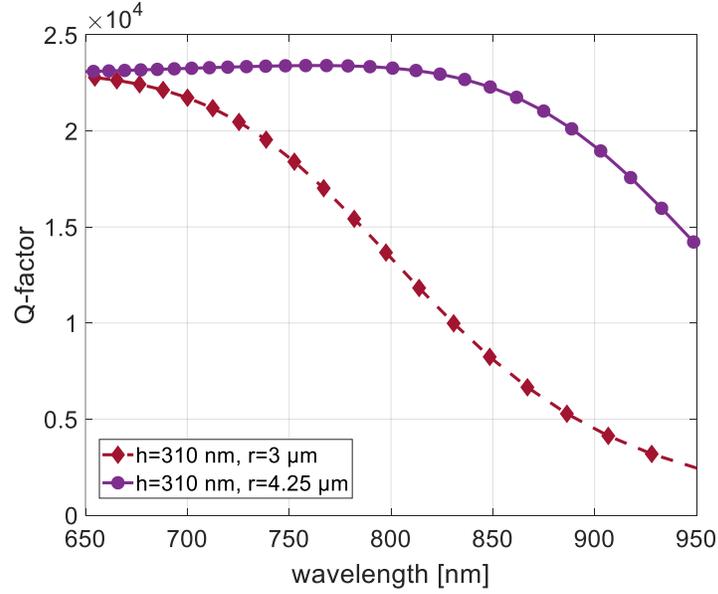

Fig. S 3. Simulated Q-factors of two h-BN disk resonators, both with a thickness of 310 nm. The radius was chosen as 3 μm (red diamonds, and red dashed line as guide for the eye) and as 4.25 μm (purple circles, and purple line as guide for the eye). For integration of the HBL with the interlayer exciton emission at 923 nm it is beneficial to increase the disk radius.

Fig. S 2c). This can be explained by the fact, that for higher wavelengths the radiative losses due to the curvature of the disk increase, leading to a decrease of the Q-factor.

## V. STUDY OF VERTICAL EMITTER POSITION INSIDE H-BN DISK RESONATOR

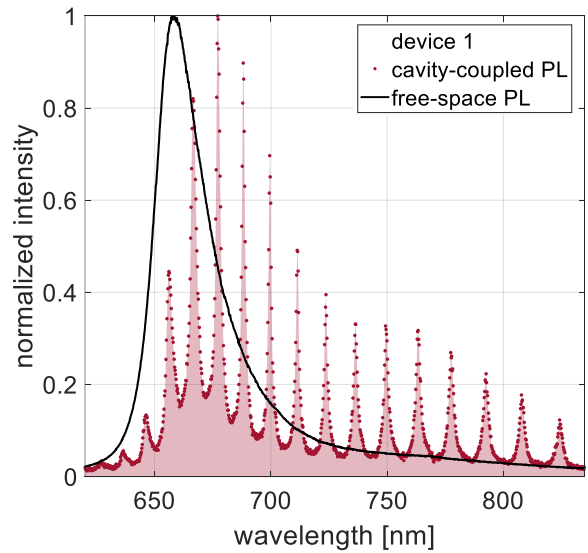

Fig. S 4. Cavity-coupled PL measurements with excitation at one waveguide facet and detection at the other waveguide end for a disk resonator with a monolayer $MoS_2$ sandwiched inside the h-BN (device 1, in Fig. S 5(c)). In addition the free-space PL measurement of the monolayer $MoS_2$ is plotted for comparison.

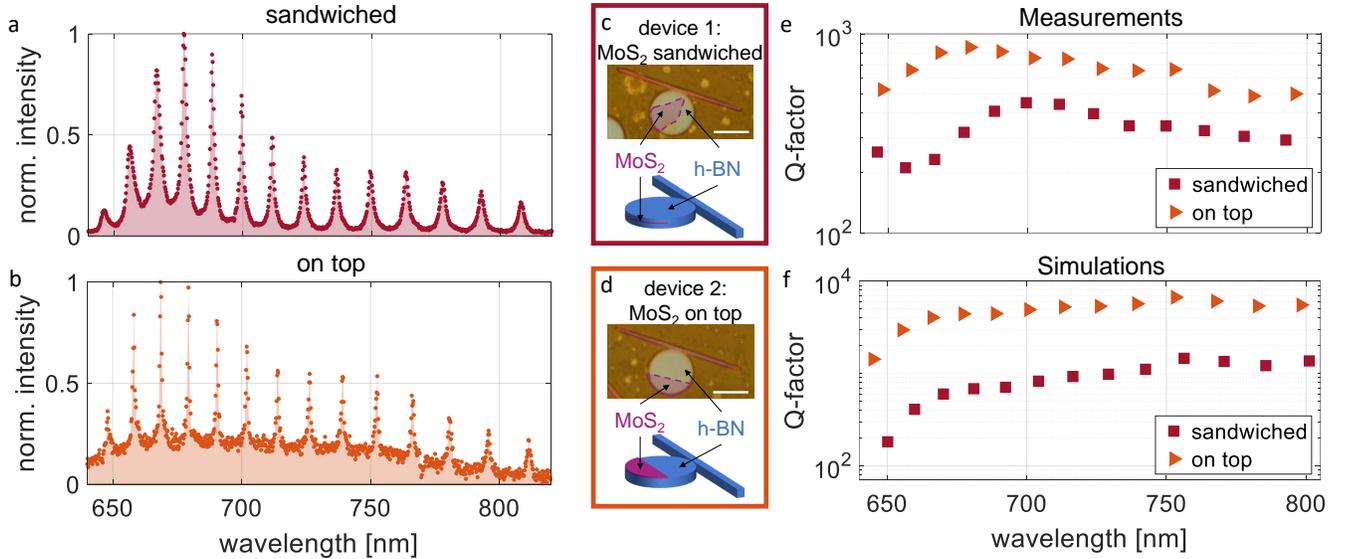

Fig. S 5. Cavity-coupled PL measurements with excitation at one waveguide facet and detection at the other waveguide end for a disk resonator with a monolayer $MoS_2$ (a) sandwiched inside the h-BN (device 1, in (c)) and (b) placed on top of the h-BN (device 2, in (d)). (c) Device 1: waveguide-coupled disk resonator with a monolayer $MoS_2$ sandwiched inside the h-BN. Corresponding data is shown in red. (d) Device 2: waveguide-coupled disk resonator with a monolayer $MoS_2$ transferred on top of the h-BN. Corresponding data is shown in orange. The scale bar in the optical microscope images in (b) and (c) is 5 µm. (e) Q-factors for device 1 (red squares) and device 2 (orange triangles), extracted from measurements. The drop at wavelengths corresponding to the intralayer exciton transition energy of the monolayer verifies the interaction of emitter and cavity. The Q-factor for device 1 is lower than for device 2, indicating the higher interaction when sandwiching the monolayer inside, placing it at higher mode intensities. (f) Simulated Q-factors for the two structures that show qualitatively similar behavior.

In the following we discuss the reabsorption by the emitter for integration at two different vertical positions. For this comparison, two different devices were fabricated, one with monolayer $MoS_2$ sandwiched inside the h-BN disk (device 1, red) and one with monolayer $MoS_2$ transferred on top of the h-BN disk (device 2, orange) (sketch and optical microscope image in Fig. S 5c and d). For device 1 the fabrication process is the same as for the structure with the integrated HBL. For device 2, an area on the same vdW stack next to device 1, but without integrated $MoS_2$ is chosen. To not expose the monolayer to chemicals during the fabrication process, the $MoS_2$ monolayer is transferred on top of the already pre-patterned h-BN–h-BN structure. Both devices have an equivalent height of 310 nm, an identical geometry (radius of 3 µm) and the same orientation.

First, we discuss the spatially resolved spectra (excitation with a cw 532 nm laser) of the cavity-coupled and the free-space PL of device 1 with $MoS_2$ sandwiched inside the disk (see Fig. S 4). For the cavity-coupled PL, the peak intensities towards energies above the intralayer exciton transition of the monolayer $MoS_2$ are reduced and the highest peak intensity is observed at wavelengths longer than the free-space PL emission peak. This is in agreement with the measurements of the device with integrated HBL, when exciting on the monolayer region (cf. Fig. 5a in the main text) and we also attribute it mainly to the reabsorption of the monolayer itself. The determined Q-factor confirms the effect of reabsorption by a strong drop for wavelengths corresponding to the intralayer exciton transition energy of the monolayer (see red squares in Fig. S 5e).

To discuss the influence of the vertical position of the emitter on the effect of reabsorption, we plot as a comparison the spectrum and the Q-factors of device 2 (spectrum in Fig. S 5b and orange triangles in e). Again a drop of the Q-factor can be observed for wavelengths around the intralayer exciton transition of $MoS_2$, verifying the reabsorption of the emitter and consequently the interaction between the cavity and the emitter that is now placed on top of the device. However, the overall Q-factor of device 2 is higher than that of device 1. This might be explained by the fact that the interaction of emitter and cavity is reduced by placing the emitter on top of the photonic structure at lower optical field intensities (cf. simulations in Fig. 3c in the main text). Consequently, the reabsorption of the cavity-coupled PL by the emitter itself is reduced. The measurements are verified, by comparison to simulations (see Fig. S 5f). The $MoS_2$ monolayer are integrated in the simulations by adding a layer with a thickness of 0.7 nm with a refractive index taken from [6]. The vertical position of the integrated layer can be adapted to the considered stack. Even though the

overall simulated Q-factor is much higher, the same trend is observable, confirming the higher interaction between cavity mode and emitter in a sandwiched structure and illustrating the resulting higher reabsorption by the emitter itself.

## VI. TRANSMISSION MEASUREMENT OF WAVEGUIDE-COUPLED DISK RESONATOR WITH INTEGRATED HBL

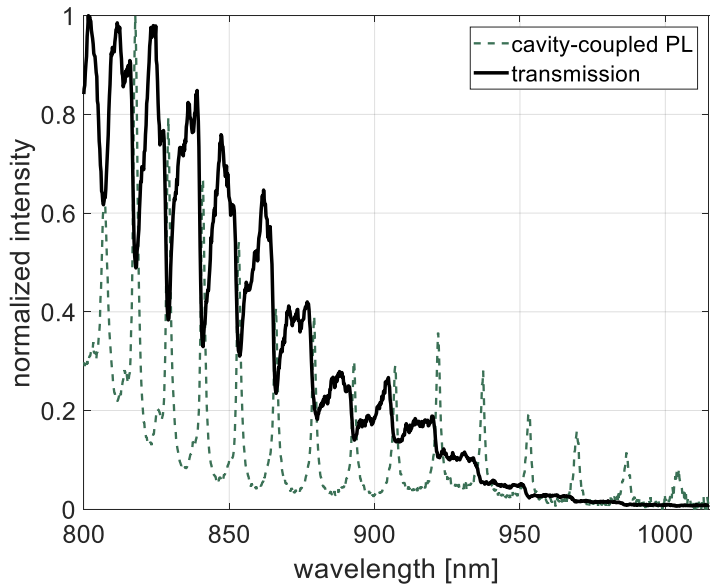

Fig. S 6. Transmission spectrum of an h-BN waveguide-coupled disk resonator with integrated heterobilayer (black line). The transmission dips agree with the resonances that the exciton emission couples to upon excitation with a red laser (dashed green line).

In Fig. S 6 (black line) the transmission measurement of a waveguide-coupled disk resonator with integrated HBL is shown. The transmission spectrum is measured with a broad band laser source, which exhibits a strong decrease in laser power for wavelengths around 900 nm and impedes measurements for this wavelength range. The measured resonance dips are in spectral agreement with the resonances the exciton emission couples to upon excitation with the red laser (dashed green line in Fig. S 6).

## VII. DISCUSSION OF THE EFFECT OF REABSORPTION INSIDE THE DISK RESONATOR

In the following we aim at developing an understanding of the optical losses to describe the envelope function of the resonances of the cavity-coupled PL (dark green curve in Fig. S 7). In our simulations we observe that the wavelength dependent Q-factor, for devices without integrated HBL, stays nearly constant within the wavelength range of 790 nm (free-space PL) and 820 nm (maximum of cavity-coupled PL). Hence, within this spectral range the cavity induced gain and losses (without absorption) should be nearly wavelength independent. In this scenario, a good way to describe the redshift is by modeling the reabsorption inside the cavity with integrated HBL. For this purpose, we used a simplified model to account for the attenuation of light inside the cavity. Re-emission by the monolayer or the HBL upon reabsorption is not considered here. The Q-factor of a disk resonator defines the photon lifetime, and thereby the average number of round trips that a photon makes inside the cavity. Consequently the photon travels a certain mean path length $s$ inside the cavity given by:

$$s = \frac{c}{n}\tau = \frac{Q}{n}\lambda, \qquad (2)$$

with $c$ being the speed of light, $n$ the refractive index, $\tau$ the photon lifetime and $Q$ the Q-factor of the disk resonator (Q $\sim$200 at $\sim$800 nm from our experiments). The reduction of the initial intensity ($I_0$) inside the disk can then be

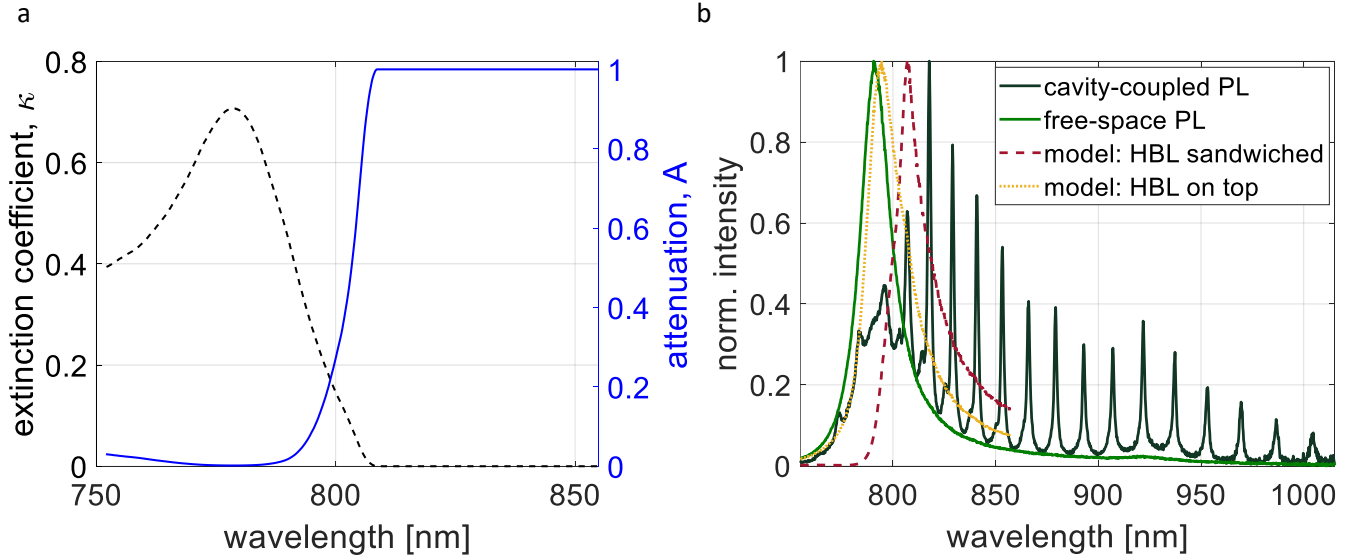

Fig. S 7. (a) Extinction coefficient of a monolayer MoSe$_2$ from [7] (dashed black line) and attenuation, A, due to the reabsorption inside the cavity (blue line). (b) Multiplying the free-space PL spectrum (light green) with the attenuation factor leads to the spectrum indicated by the dashed red line. Similar to the measured cavity-coupled PL (dark green curve), a shift of the maximum peak towards longer wavelengths can be observed. For the dark red curve the values are adjusted such that the optical mode overlap corresponds to an emitter sandwiched in between two h-BN flakes. Whereas for the dotted yellow curve the values are adjusted such that the optical mode overlap corresponds to an emitter placed on top of the h-BN. A reduced shift towards higher wavelengths can be observed, due to a reduced mode overlap with the emitter and consequently a reduced reabsorption.

determined using the Lambert-Beer law [8].

$$I = I_0 \exp\left(-\alpha s\right) = I_0 \exp\left(-2\kappa(\lambda)\frac{2\pi}{\lambda}s\right) = I_0 \underbrace{\exp\left(-4\pi\kappa(\lambda)\frac{Q}{n}\right)}_{A} \qquad (3)$$

where $\alpha$ is the absorption coefficient, $\kappa(\lambda)$ is the extinction coefficient, i.e. the imaginary part of the refractive index for a wavelength $\lambda$.

Owing to the small thickness of the monolayer MoSe$_2$, and considering that it only covers half of the area of the disk resonator, only a fraction of the mode overlaps with the MoSe$_2$. From numerical simulations we find an optical confinement factor $\Gamma$ [9] of around 1.5 %. Figure S 7 shows the effect of the reabsorption inside the cavity: The black dashed line in Fig. S 7a shows the extinction coefficient, $\kappa(\lambda)$ that has been taken from [7]. The blue line in Fig. S 7a shows the attenuation, A, caused by MoSe$_2$, after travelling a mean path length $s$ in the disk resonator. Multiplying the free-space PL spectrum (light green curve in Fig. S 7b) of MoSe$_2$ with the attenuation leads to the dashed dark red curve in Fig. S 7b. A shift towards longer wavelengths is apparent. The maximum peak of this modelled spectrum is strongly redshifted towards the peak of our experimental cavity-coupled PL spectrum (dark green curve). The discrepancy between the modeled envelope function and the measured cavity-coupled PL, might be explained by variations of $\kappa(\lambda)$ of our sample to the used value from [7]. The magnitude of the redshift depends on the mode overlap, which is enhanced in our sandwiched structure. In devices where the monolayer is placed on top of the disk a smaller reabsorption is expected. The reduced redshift can also be visualized by the attenuated spectrum for a reduced optical mode overlap (dotted yellow line in Fig. S 7b, from simulations we observe a decrease of $\Gamma$ to 0.3 %).

While this model explains most of the redshift, other effects might contribute too. For example, a recent study has reported phonon-exciton interaction to cause redshifting [10]. In this study, however, the monolayer was placed on top of the photonic structure, hence it has a reduced mode overlap and therefore smaller reabsorption.

## VIII. DETERMINATION OF Q-FACTOR

To determine the wavelength dependent Q-factor (plotted in Fig. 5b in the main text), each resonance of the measured spectra was fitted, as shown in the example spectra and the corresponding fits in Fig. S 8a. The individual peaks were fitted with Lorentzian functions and a linear background (magnified for a representative resonance peak in Fig. S 8b). To minimize the influence of the background, we calculate the mean value and the standard deviation for each fitted Q-factor for different excitation positions (at both waveguide ends, on the monolayer region and on the interlayer region), as well as for detection at both waveguide ends separately. In addition, we also consider the fitted Q-factors for a grating with 600 grooves per mm and a slit opening of 20 µm, with a better resolution, but in return a reduced intensity.

## IX. COMPARISON OF THE RESONANCE PEAK INTENSITIES FOR THE WAVEGUIDE-COUPLED DISK RESONATOR WITH INTGRATED HETEROBILAYER

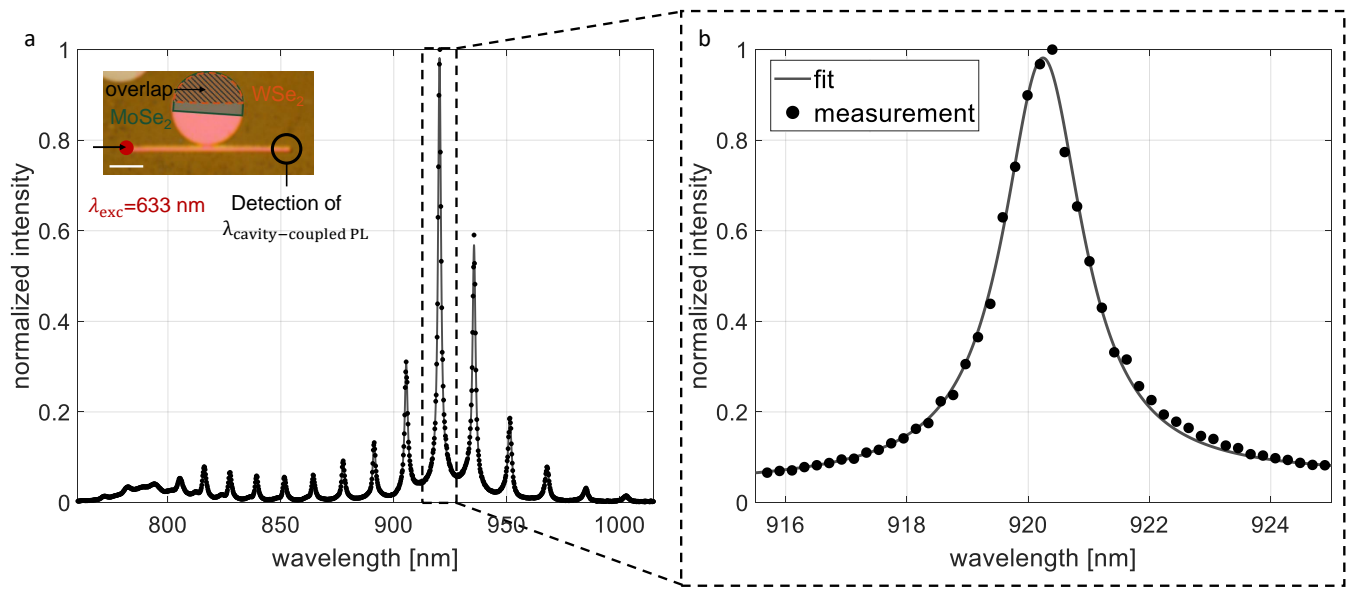

Fig. S 8. (a) Measured data when exciting a device with integrated HBL at one waveguide end and detecting at the other, as shown in the inset. The fit for each resonance is overlaid. The scale bar in the inset is 5 µm. (b) Magnification of one resonance. Each resonance is fitted separately by a Lorentzian function and a linear background.

To discuss the direct absolute comparison of the intralyer and the interlayer exciton coupling to the whispering gallery modes of the h-BN disk resonator with integrated HBL, we consider the spectrum detected at one waveguide end, when exciting at the other waveguide facet (Fig. S 8a, same as Fig. 1c in the main text). Note that the spectrum is corrected for the transmission of the experimental setup and the detection efficiency of the CCD. We can evaluate the difference between the intensity of the two main emission peaks (∼920 nm and ∼820 nm) of the interlayer and intralayer peaks either by comparing the maximum peak intensity or by integration over the fit of the peak. The coupled interlayer vs. intralayer emission ratio is then determined to be 12.8 and 12.9, respectively for the shown measurement with an excitation power of ∼186 µW before the objective.

However, it is important to point out that the interaction length of the cavity mode and the HBL region compared to the monolayer region is different. The interaction length within the HBL region is 4 times longer than within the monolayer region, when estimating the ratio via the circumference of the disk. It is remarkable, that even after considering this factor, the interlayer exction emission in the cavity-coupled PL measurements is higher. Note that the wavelength dependent efficiencies for the out-coupling from the WGMs inside the disk to the waveguide facet

could also influence these measurements.



\end{document}